\documentclass[preprint,aps ,nofootinbib]{revtex4}
\usepackage{graphicx}
\pdfoutput=1
\usepackage{epsfig}
\usepackage{amsmath}
\usepackage{amsfonts}
\usepackage{amssymb}
\usepackage{color}
\usepackage{dcolumn}
\usepackage{hyperref}
\usepackage{multirow}
\usepackage{booktabs}
\providecommand{\U}[1]{\protect\rule{.1in}{.1in}}

\newcommand{\f}{\begin{equation}}
\newcommand{\ff}{\end{equation}}
\newcommand{\fa}{\begin{eqnarray}}
\newcommand{\ffa}{\end{eqnarray}}

\begin{document}
\title{The shadow of regular black holes with asymptotically Minkowski core }
\author{Yi Ling $^{1,2}$}
\email{lingy@ihep.ac.cn}
\author{Meng-He Wu$^{3,4}$}
\email{mhwu@sues.edu.cn} \affiliation{$^1$Institute of High Energy
Physics, Chinese Academy of Sciences, Beijing 100049, China\\ $^2$
School of Physics, University of Chinese Academy of Sciences,
Beijing 100049, China \\
$^3$ School of Mathematics, Physics and Statistics, Shanghai University of Engineering Science, Shanghai 201620, China \\
$^4$Center of Application and Research of Computational Physics, Shanghai University of Engineering Science, Shanghai 201620, China}

\begin{abstract}
We investigate the shadow cast by a sort of new regular black holes which are characterized by an asymptotically Minkowski core and the Sub-Planckian curvature. Firstly, we extend the metric with spherical symmetry to the one of rotating Kerr-like black holes and derive the null geodesics with circular orbit near the horizon of the black hole, and then plot the shadow of black holes with different values of the deviation parameter. It is found that the size of the shadow shrinks with the increase of the deviation parameter, while the shape of the shadow becomes more deformed. In particular, by comparing with the shadow of Bardeen black hole and Hayward black hole with the same values of parameters, we find that in general the shadow of black holes with Minkowski core has a larger deformation than that with de Sitter core, which potentially provides a strategy to distinguish these two sorts of regular black holes with different cores by astronomical observation in future.

\end{abstract}
\maketitle
\section{Introduction}

The release of the first photo of black hole shadows by the Event Horizon Telescope (ETH) cooperation team has announced the coming of a new age for measuring the nature of astrophysical black holes \cite{EventHorizonTelescope:2019dse,EventHorizonTelescope:2019ths}.
Just recently, the ETH collaboration released a photo of the supermassive black hole Sgr A*, which provides overwhelming evidence for the existence of black hole at the center of the Milky Way\cite{Akiyama2022L12,Akiyama2022L14}.
Now it is fair to say that more and more astronomical evidences have been making it possible for people to justify various theoretical thoughts on black holes through the experimental observation of black hole shadows. As a matter of fact,  the shadow of different black holes has previously been explored theoretically in \cite{Synge:1966okc,Luminet:1979nyg,Bardeen:1973,Chandrasekhar:1992,Cunha:2018acu,Hu:2020usx,Ling:2021vgk}, and a recent review on this topic can be found in \cite{Perlick:2021aok}. For instance, the shadow of a Schwarzschild black hole was firstly studied by Synge in \cite{Synge:1966okc} and Luminet in \cite{Luminet:1979nyg}. Without surprise, they found that its shadow was a perfect circle. A little bit of later, the shadow of the Kerr black hole was studied in \cite{Bardeen:1973, Chandrasekhar:1992}. Its shadow is no longer circular, but has a deformation in the direction of rotation, which provides an abundant structure for the shadow of black holes and in principle one may obtain the nature of rotating black holes by observing its shadow, such as the mass, spin and charge of black holes \cite{Takahashi:2005hy,Hioki:2009na,Tsupko:2017rdo,Bambi:2019tjh,Badia:2021kpk}. By virtue of black hole shadows, some fundamental problems on the nature of gravity have also been explored, such as the modified gravity, the candidates for dark matter, and the quantum effect of gravity, etc. \cite{Johannsen:2010ru,Amir:2016cen,Xu:2018mkl,Guo:2020zmf,Liu:2020ola,Wei:2013kza,Yan:2019hxx,Badia:2020pnh,Vagnozzi:2019apd,Khodadi:2020jij,Vagnozzi:2022moj,Lee:2021sws}.

Recently, the investigation on the shadows has been extended to
regular black holes as well, including rotating Bardeen and
Hayward black
holes\cite{Abdujabbarov:2016hnw,Tsukamoto:2017fxq,Kumar:2019pjp,Kumar:2020xvu}.
In history, regular black holes are proposed to resolve the
singularity problem in classical general relativity. Taking the
quantum effects of gravity into account, it is believed that these
singularities could be removed or avoided. Before a well-defined
theory of quantum gravity could be established, people have
constructed various regular black holes without singularity at the
phenomenological level
\cite{Bardeen:1968hnw,Hayward:2005gi,Frolov:2014jva,Ansoldi:2008jw,Xiang:2013sza,Ling:2021olm,Ling:2021abn}.
These regular black hole models are characterized by a finite
Kretschmann scalar curvature.
Currently, as far as we know, there exist various ways
to classify the regular black holes either by their symmetry\cite{Bargueno:2020ais,Melgarejo:2020mso} or by their asymptotic
behavior near the center of the black hole. In this paper, we
classify the regular black holes into two classes according to the
latter.
One is the regular black holes with asymptotically de
Sitter core, such as Bardeen black hole, Hayward black hole as
well as  Frolov black hole. The other one is the regular black
holes with asymptotically Minkowski core, which are characterized
by an exponentially suppressing Newton potential.

Our current work is motivated to answer the following
question: How to diagnose a detected black hole to be an ordinary
black hole with singularity rather than a regular black hole
without singularity? Without doubt this is a very crucial issue on
the nature of detected black holes and has significant impacts on
the theory of gravity. Unfortunately, since the singularity hides
inside the horizon, currently it is impractical to distinguish
these two kinds of black holes by diving into the black hole or
detecting any signal coming out from the interior of the horizon.
Nevertheless, with the accumulation of the observation data on
detected black holes, we are wondering if one can distinguish a
regular black hole from a traditional black hole with singularity
by observing the shadow of black holes. Therefore, investigating
the shadow of regular black holes theoretically just as performed
in this manuscript could shed light on this issue and improve our
understanding on the observation data of detected black holes.
Previously, the shadow of regular black holes with de Sitter
core has been investigated in \cite{Abdujabbarov:2016hnw},
 and the study of shadows
has also been extended to regular black holes of the Minkowski
core\cite{Banerjee:2022iok,Simpson:2021dyo,Berry:2020tky}.
 However the black holes with Minkowski core
previously studied in literature have a shortcoming. From
the viewpoint of quantum gravity, the Kretschmann scalar curvature
of a regular black hole should be  sub-Planckian
everywhere and anytime, whereas the Kretschmann scalar curvature
of the previously discussed black hole with Minkowski core is
mass dependent such that its Kretschmann scalar curvature could
easily exceed the Planck mass density by increasing the mass of
the black hole. This implies that the regular black holes with
Minkowski core previously used to study black hole shadows only
make sense for small masses at the Planck scale \cite{Ling:2021olm}. On the
other hand, the Kretschmann scalar curvature of Bardeen black hole
and Hayward black holes which have de-Sitter core is mass
independent\cite{Ling:2021olm}. The scalar curvature of this sorts of black holes exhibits distinct behavior.
Thus in previous literature no one could
compare the shadow of these two sorts of black holes. In this
paper, we intend to investigate the shadow of the regular black
hole with Minkowski core characterized by mass independent Kretschmann scalar
curvature\cite{Ling:2021olm} and compare it with the shadow of the one with the same
parameter values but with de-Sitter core.

We organize the paper as follows. In next section we extend the
regular black hole with spherical symmetry proposed in \cite
{Ling:2021olm} to the rotating Kerr-like one, and then obtain the
null geodesics with circular orbit near the horizon of black hole
with the Hamilton-Jacobi formalism. In section three and four, we
will investigate the shadow of regular black holes with
$(\gamma=2/3, n=2)$ and $(\gamma=1,n=3)$, respectively, and
compare them with the shadow of Bardeen and Hayward black holes
with the same values of parameters. Furthermore, we will
compute the upper limit of the deviation from circularity and
compare our theoretical results with the observation data on the
shadow of supermassive black hole at the center of the galaxy M87*
by ETH. It is found that the shadow of regular black holes is well
compatible with the observed shadow by ETH as well. Our
conclusions and discussions are given in section five.

\section{Rotating Kerr-Like regular black hole with Minkowski core}
In \cite {Ling:2021olm} we have proposed a new sort of regular spherically symmetric black holes with the following metric
\begin{equation}
\begin{aligned}
d s^{2} =-f(r) d t^{2}+f(r)^{-1} d r^{2}
+r^2( d \theta^{2}+ \sin ^{2} \theta d \phi^{2}),
\end{aligned}\label{msm}
\end{equation}
with
\begin{equation}
\begin{aligned}
f(r) =1-\frac{2 m(r) }{r},
\end{aligned}
\end{equation}
where $m(r)$ takes the form as
\begin{equation}
m(r)=M e^{-g^n M^{\gamma }/ r^{n}}.\label{mm}
\end{equation}
We have set $G=1$ throughout this paper.
We remark that the above metric form can be understood
as the solution of Einstein field equation in which the
gravitational field is coupled to a nonlinear Maxwell field. The
physical source of the regular black hole could be interpreted as
the nonlinear electromagnetic field \cite{Ayon-Beato:1998hmi}. The generic form of the
stress-energy tensor and the discussion on the violation of the
strong energy condition is presented in \cite {Ling:2021olm}.

This sort of regular black holes exhibits  the following two prominent characters. Firstly, the exponentially suppressing form of Newton potential leads to a non-singular Minkowski core at the center of the black hole, which was originally proposed in \cite{Xiang:2013sza}, but with the specific form of $\gamma=0, n=2$. Secondly, under the condition $3/n<\gamma<n$, the Kretschmann scalar curvature can be always sub-Planckian regardless the mass of black holes once the parameter $g$ is appropriately fixed. Finally, the correspondence of such regular black holes to the ones with asymptotically de Sitter core has been pointed out in  \cite {Ling:2021olm}, where $m(r)$ takes the form as
\begin{equation}
m(r)=\frac{M r^{\frac{n}{\gamma}}}{(r^n+\gamma g^n M^{\gamma})^{1/\gamma}}.
\end{equation}
Specially, $\gamma=2/3,n=2$ leads to Bardeen black hole, while $\gamma=1,n=3$  leads to Hayward black hole.

In this paper in order to obtain a non-circular shadow with distortions, we extend the above metric with $m(r)$ given in (\ref{mm}) to describe the rotating Kerr-like black hole.
We employ the Newman-Janis algorithm\cite{Newman:1965,Azreg-Ainou:2014pra,Azreg-Ainou:2014aqa,wu:2022aqa,Ghosh:2014pba,Azreg-Ainou:2014nra,Toshmatov:2014nya}  for the static spherical regular black hole (\ref{msm}) and obtain the rotating  regular black hole.
For a spherical symmetric metric (\ref{msm}),
we introduce $d u=d t-d r / f(r)$ to get null coordinates $\{u, r, \theta, \phi\}$. The metric  becomes
\begin{equation}\label{Eq_symq2}
d s^2=-f(r) d u^2-2  d u d r+r^2\left(d \theta^2+\sin ^2 \theta d \phi^2\right).
\end{equation}
The inverse metric $g^{\mu \nu}$ can be represented by a null tetrad$ \left(l^\mu, n^\mu, m^\mu, m^{\dagger \mu}\right)$ as
\begin{equation}\label{Eq_symq3}
g^{\mu \nu}=-l^{\mu^\nu} n^v-l^v n^\mu+m^\mu m^{\dagger \nu}+m^v m^{\dagger \mu},
\end{equation}
with the relation
\begin{equation}\label{Eq_symq4}
\begin{aligned}
&l_\mu l^\mu=n_\mu n^\mu=m_\mu m^\mu=l_\mu m^\mu=n_\mu m^\mu=0, \ l_\mu n^\mu=-m_\mu m^{\dagger \mu}=-1.
\end{aligned}
\end{equation}
Then the null tetrad $\left(l^\mu, n^\mu, m^\mu, m^{\dagger \mu}\right)$ can be expressed as
\begin{equation}\label{Eq_symq5}
l^\mu=\delta_r^\mu, \quad n^\mu=\delta_u^\mu-\frac{f(r)}{2} \delta_r^\mu, \quad m^\mu=\frac{1}{\sqrt{2 r^2}}\left(\delta_\theta^\mu+\frac{i}{\sin \theta} \delta_\phi^\mu\right).
\end{equation}
With the following complex coordinate transformation
\begin{equation}\label{Eq_symq6}
 r^{\prime}=r+i a \cos \theta, \quad u^{\prime}=u-i a \cos \theta,
\end{equation}
the null tetrad $\left(l^\mu, n^\mu, m^\mu, m^{\dagger \mu}\right)$ becomes
\begin{equation}\label{Eq_symq7}
\begin{aligned}
&\quad \quad  \quad  l^{\prime \mu}=\delta_r^\mu, \quad n^{\prime \mu}=\delta_u^\mu-\frac{\tilde{f}(r)}{2} \delta_r^\mu, \\
 m^{\prime \mu}=&\frac{1}{\sqrt{2\left(r^{\prime}-i a \cos \theta\right)^2}}\left(i a \sin \theta\left(\delta_u^\mu-\delta_r^\mu\right)+\delta_\theta^\mu+\frac{i}{\sin \theta} \delta_\phi^\mu\right),
\end{aligned}
\end{equation}
with
 \begin{equation}\label{Eq_symq8}
\tilde{f}(r)=1-\frac{2 m(r) r}{\Sigma},
\end{equation}
where $\Sigma=r^{2}+a^{2} \cos ^{2} \theta$
and $a$ is denoted as the spin parameter of an axisymmetric black hole. Using  Eq.(\ref{Eq_symq7}), we can get a new inverse metric $g'^{\mu \nu}$ :
\begin{equation}\label{Eq_symq9}
g^{\prime \mu v}=-l^{\prime \mu} n^{\prime \nu}-l^{\prime \nu} n^{\prime \mu}+m^{\prime \mu} m^{\prime \dagger \nu}+m^{\prime \nu} m^{\prime \dagger \mu},
\end{equation}
whose explicit metric formula is
\begin{equation}\label{Eq_symq10}
\begin{aligned}
d s^2 &=-\tilde{f}(r) d u^2-2 d u r^{\prime}+2 a \sin ^2 \theta\left(\tilde{f}(r)-1\right) d u d \phi+2 a \sin ^2 \theta d r d \phi \\
&+\left(r^{\prime}-i a \cos \theta\right)^2 d \theta^2+\sin ^2 \theta\left[\left(r^{\prime}-i a \cos \theta\right)^2+a^2 \sin ^2 \theta\left(2 -\tilde{f}(r)\right)\right] d \phi^2.
\end{aligned}
\end{equation}
Finally we apply the following transformations to obtain the metric in Boyer-Lindquist coordinates
\begin{equation}\label{Eq_symq11}
d u=d t^{\prime}-\frac{\Sigma+a^2 \sin ^2 \theta}{\Sigma f\left(r\right)+a^2 \sin ^2 \theta} d r, \quad d \phi=d \phi^{\prime}-\frac{a}{\Sigma f\left(r\right)+a^2 \sin ^2 \theta} d r,
\end{equation}
where $\Sigma=r^{2}+a^{2} \cos ^{2} \theta$.
The new metric is given by
\begin{equation}\label{Eq_symq12}
\begin{aligned}
d s^2=&-\tilde{f}(r) d t^2+\frac{\Sigma}{\Sigma \tilde{f}(r)+a^2 \sin ^2 \theta} d r^2 -2 a \sin ^2 \theta(1-\tilde{f}(r)) d \phi d t\\
&+\Sigma d \theta^2 +\sin ^2 \theta\left[\Sigma-a^2(\tilde{f}(r)-2) \sin ^2 \theta\right] d \phi^2.
\end{aligned}
\end{equation}
Substituting (\ref{Eq_symq8}) into (\ref{Eq_symq12}), we obtain the rotating regular black hole in Boyer-Lindquist coordinates
\begin{equation}\label{Eq_BL}
\begin{aligned}
d s^{2} &=-\left(1-\frac{2 m(r) r}{\Sigma}\right) d t^{2}-\frac{4 a m(r) r \sin ^{2} \theta}{\Sigma} d t d \phi+\frac{\Sigma}{\Delta} d r^{2} \\
&+\Sigma d \theta^{2}+\left(r^{2}+a^{2}+\frac{2 a^{2} m(r) r \sin ^{2} \theta}{\Sigma}\right) \sin ^{2} \theta d \phi^{2},
\end{aligned}
\end{equation}
with
\begin{equation}
 \quad \Delta=r^{2}-2 m(r) r+a^{2}.
\end{equation}
In this Kerr-like metric, $a$ is the rotation parameter and obviously as $a\rightarrow 0$ it goes back to the metric given in (\ref{msm}).Now, we plot the shadows of these sort of black holes closely following the route presented in \cite{Abdujabbarov:2016hnw}, where the shadow of black holes with different mass functions $m(r)$ has been investigated for Bardeen and Hayward black holes.

Firstly, we consider the null geodesics near the horizon of the black hole. We start with a Lagrangian system for a photon
\begin{equation}
\mathcal{L}=\frac{1}{2}\left(\frac{d s}{d \sigma}\right)^{2}=\frac{1}{2} g_{\mu \nu} \dot{x}^{\mu} \dot{x}^{\nu},
\end{equation}
where $\sigma$ is an affine parameter along the geodesics.
Over a Kerr-like black hole background,  $\partial /\partial t$ and $\partial/\partial \phi$ are killing vectors, thus there are two conservative quantities, corresponding to  the energy and the angular momentum of the photon, respectively:
\begin{equation}
\begin{aligned}
E&=-p_{t}= g_{tt} \dot{t}, \\
L_{z}&=p_{\phi}= g_{\phi \phi} \dot{\phi}.
\end{aligned}
\end{equation}
We can easily get:
\begin{equation}
\begin{aligned}
\Sigma \frac{d t}{d \sigma}&=a\left(L_{z}-a E \sin ^{2} \theta \right)+\frac{r^{2}+a^{2}}{\Delta}\left[\left(r^{2}+a^{2}\right) E-a L_{z}\right], \\
\Sigma \frac{d \phi}{d \sigma}&=\left(\frac{L_{z}}{\sin ^{2} \theta}-a E\right)+\frac{a}{\Delta}\left(\left(r^{2}+a^{2}\right) E-a L_{z}\right).
\end{aligned}
\end{equation}

 The Hamilton-Jacobi equation for a null geodesics is given
by
\begin{equation}
\frac{\partial S}{\partial \sigma}=-\mathcal{H},
\end{equation}
where $S$ is Jacobi action and the Hamilton for a photon has the
form
\begin{equation}\label{Eq_Ham}
\mathcal{H}=p_\mu \dot{x^\mu}-\mathcal{L}=\frac{1}{2} g^{\mu \nu} p_\mu p_\nu=0,
\end{equation}
with $p_\mu \equiv \frac{\partial S}{\partial x^\mu}$ being the
conjugate momentum of the photon. The system we consider is an
integrable system, thus the action can be expressed as:
\begin{equation} \label{Eq_Joc}
S=-E t+L_z \phi+S_{r}(r)+S_{\theta}(\theta).
\end{equation}
Substituting Eq. \ref{Eq_BL} and \ref{Eq_Joc} into Eq.\ref{Eq_Ham}, we obtain
\begin{equation}\label{Eq_rt}
\begin{aligned}
\left(\frac{d S_\theta}{d \theta}\right)^2 &+\Delta\left(\frac{d S_r}{d r}\right)^2-\left(\frac{1}{\Delta}\left(r^2+a^2\right)^2-a^2 \sin ^2 \theta\right) E^2 \\
&+\frac{4 a  m(r)r}{\Delta} E L+L^2\left(\frac{1}{\sin ^2 \theta}-\frac{a^2}{\Delta}\right)=0.
\end{aligned}
\end{equation}
Due to the fact that coordinates $r$ and $\theta$ are separable,
we can rewrite  Eq. \ref{Eq_rt} as
\begin{equation}
\begin{aligned}
&\Delta\left(\frac{d S_r}{d r}\right)^2-\frac{1}{\Delta}\left(r^2+a^2\right)^2 E^2+\frac{4 a r m(r)}{\Delta} E L-\frac{a^2}{\Delta} L^2 \\
=&-\left(\frac{d S_\theta}{d \theta}\right)^2-a^2 E^2 \sin ^2 \theta-\frac{L^2}{\sin ^2 \theta}= \mathcal{K},
\end{aligned}
\end{equation}
where $\mathcal{K}$ is the Carter constant. The relationship between $x^{\mu}$ and $p^{\mu}$ is
\begin{equation}
\frac{d x^\mu}{d \sigma}=p^\mu.
\end{equation}
Thus the geodesic equation of motion for $r$ and $\theta$ is given by
\begin{equation}
\begin{aligned}
\Sigma \frac{d r}{d \sigma}=\pm \sqrt{\mathcal{R}}, \\
\Sigma \frac{d \theta}{d \sigma}=\pm \sqrt{\Theta},
\end{aligned}
\end{equation}
with
\begin{equation}
\begin{gathered}
\mathcal{R}=\left[\left(r^{2}+a^{2}\right) E-a L_{z}\right]^{2}-\Delta\left[\mathcal{K}+\left(L_{z}-a E\right)^{2}\right], \\
\Theta=\mathcal{K}+\cos ^{2} \theta\left(a^{2} E^{2}-\frac{L_{z}^{2}}{\sin ^{2} \theta}\right),
\end{gathered}
\end{equation}

The critical condition for unstable circular orbits is given by $\mathcal{R}(r)=0$ and $d \mathcal{R}(r) / d r=0$. And we introduce $\xi=L_{z} / E$ and $\eta=\mathcal{K} / E^{2}$ to determine unstable circular orbits. We can easily get the parameters $\xi$ and $\eta$ as:
\begin{equation}
\begin{aligned}
\xi =&\frac{\left(a^2-3 r^2\right) m+r \left(a^2+r^2\right)
   \left(1+m'\right)}{a \left(m+r \left(-1+m'\right)\right)},  \\
\eta =&-\frac{r^3 }{a^2}(r^3+9 r m^2+2 \left(2 a^2 r+r^3\right) m'+r^3 m'^2 \\
&- 2 m (2 a^2+3 r^2+3 r^2 m'))  \left(m+r
   \left(-1+m'\right)\right)^{-2},
\end{aligned}
\end{equation}
where $m$ is the function of $r$ and the prime denotes the derivative with respect to the radius $r$.

Following \cite{Abdujabbarov:2016hnw},  we introduce celestial coordinates $(\alpha, \beta)$ to visualize the black hole shadow:
\begin{equation}
 \begin{aligned}
\alpha&=\lim _{r_0 \rightarrow \infty}\left(-r_{0}^{2} \sin \theta_{0} \frac{d \phi}{d r}\right), \\
\beta&=\lim _{r_{0} \rightarrow \infty}\left(r_{0}^{2} \frac{d \theta}{d r}\right),
\end{aligned}
\end{equation}
where $r_0$ is the distance between the black hole and the observer, and $\theta_0$ is the inclination angle between the rotating axis of the black hole and the observer's line of sight.
For the limit $r\rightarrow \infty$, the  celestial coordinates have the following simple form:
\begin{equation}
\begin{aligned}
\alpha&=-\xi \csc \theta_{0}, \\
\beta&=\pm \sqrt{\eta+a^{2} \cos ^{2} \theta_{0}-\xi^{2} \cot ^{2} \theta_{0}}.
\end{aligned}
\end{equation}

\begin{figure} [th]
  \center{
  \includegraphics[scale=0.3]{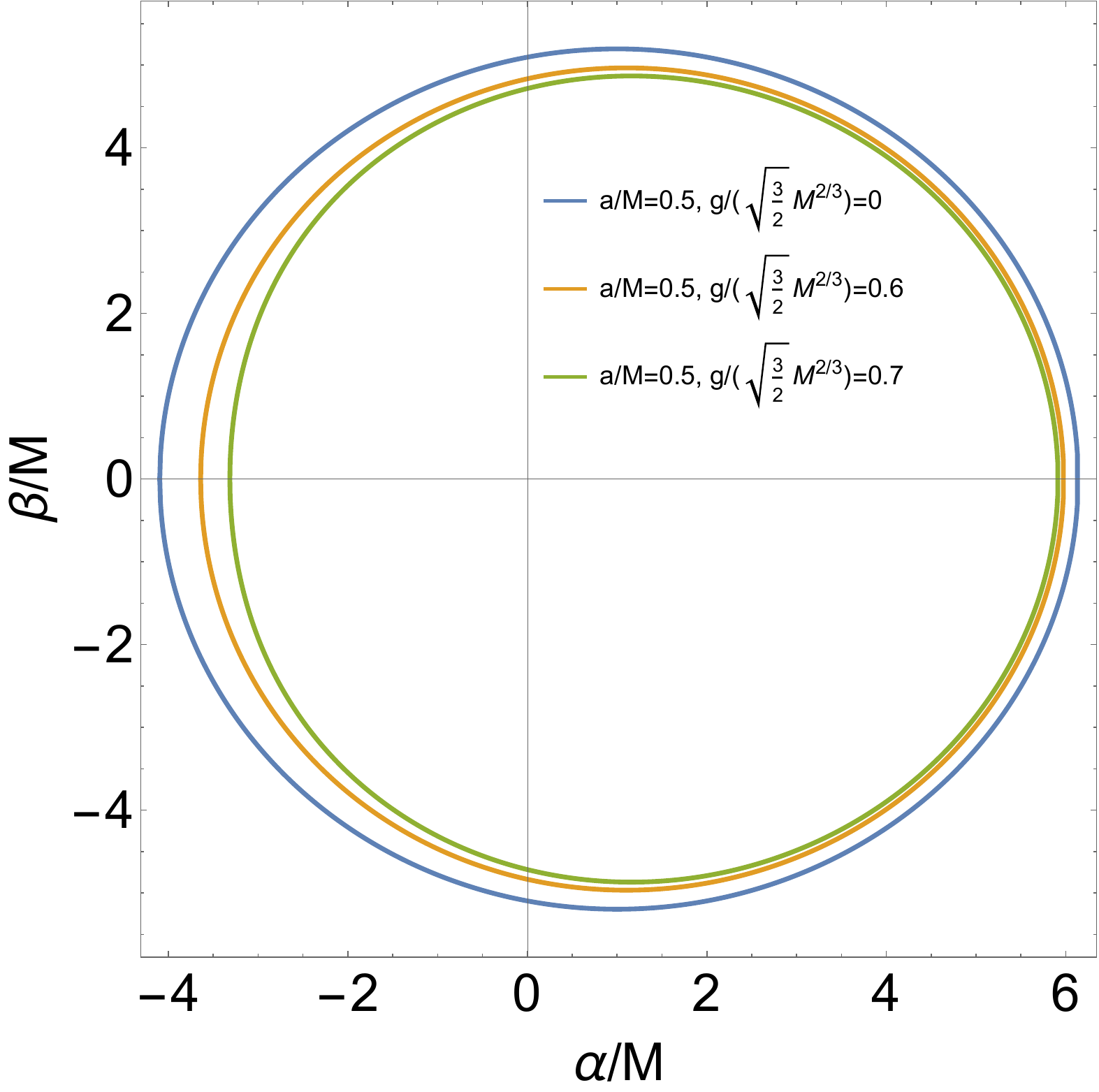}\ \hspace{0.05cm}
  \includegraphics[scale=0.3]{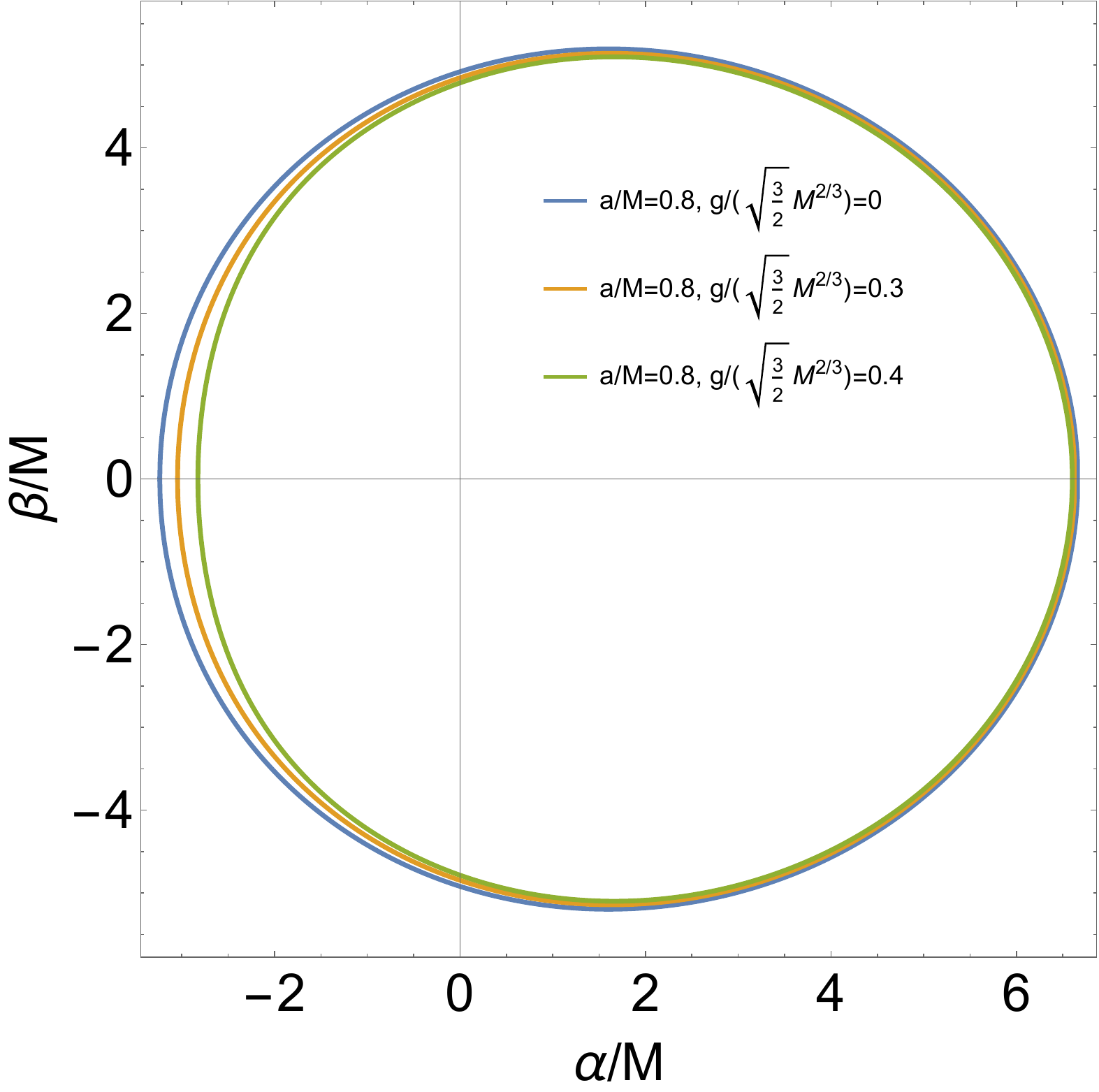}\ \hspace{0.05cm}
  \caption{The silhouette of the shadow cast by the regular black hole with  $\gamma=2/3$ and $n=2$ for various values of $g /\left(\sqrt{\frac{3}{2}} M^{2 / 3}\right)$  with $\theta_0=\pi/2$. For the left plot, the rotation parameter $a/M$ is fixed as $a/M=0.5$ while for the right plot it is fixed as $a/M=0.8$.}\label{fig_1}
  }
\end{figure}

\section{Shadow of The Regular Black Hole with $\gamma=2/3$ and $n=2$}

In this section, we study the shadow cast by the regular black hole with $\gamma=2/3$ and $n=2$, which exhibits the same asymptotical behavior as Bardeen black hole at large scale in radial direction.

In Fig.\ref{fig_1}, we plot the silhouette of the shadow for various values of the rotation parameter $a/M$ and the deviation parameter $g /\left(\sqrt{\frac{3}{2}} M^{2 / 3}\right)$, which are understood as dimensionless quantities. Firstly, by comparing the left plot with the right one, we find that  with the increase of the rotation  parameter $a/M$, the left side of the silhouette of shadow is more inclined to the vertical axis, which is similar as the phenomenon observed in \cite{Abdujabbarov:2016hnw}. Secondly, from both plots in  Fig.\ref{fig_1}  we notice that for the same rotation parameter $a/M$,  the size of the shadow shrinks with the increase of the deviation parameter $g /\left(\sqrt{\frac{3}{2}} M^{2 / 3}\right)$. Moreover, the silhouette of shadow is more deformed for the larger values of $a/M$ and $g /\left(\sqrt{\frac{3}{2}} M^{2 / 3}\right)$.

\begin{figure} [th]
  \center{
  \includegraphics[scale=0.3]{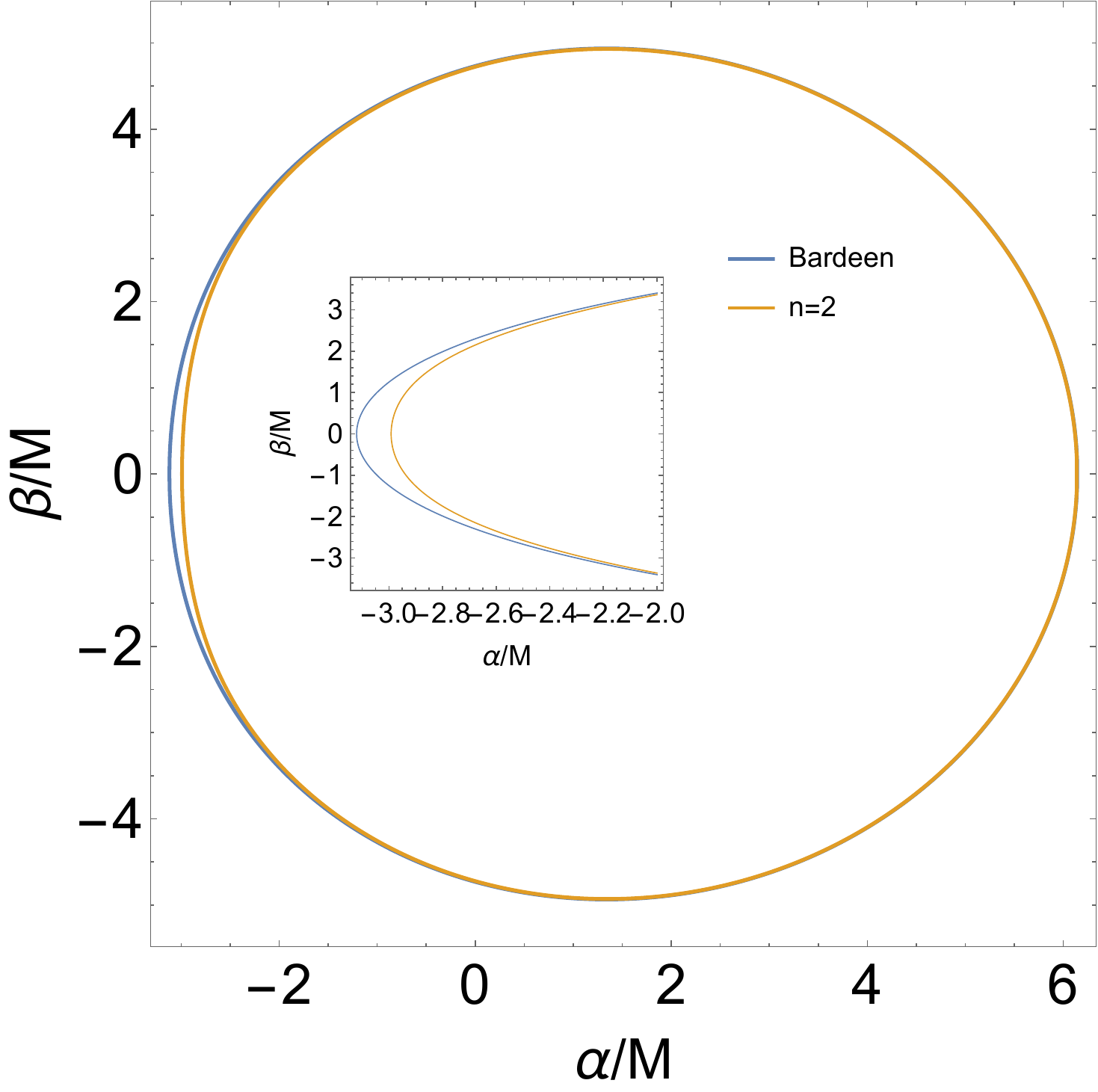}\ \hspace{0.05cm}
  \includegraphics[scale=0.3]{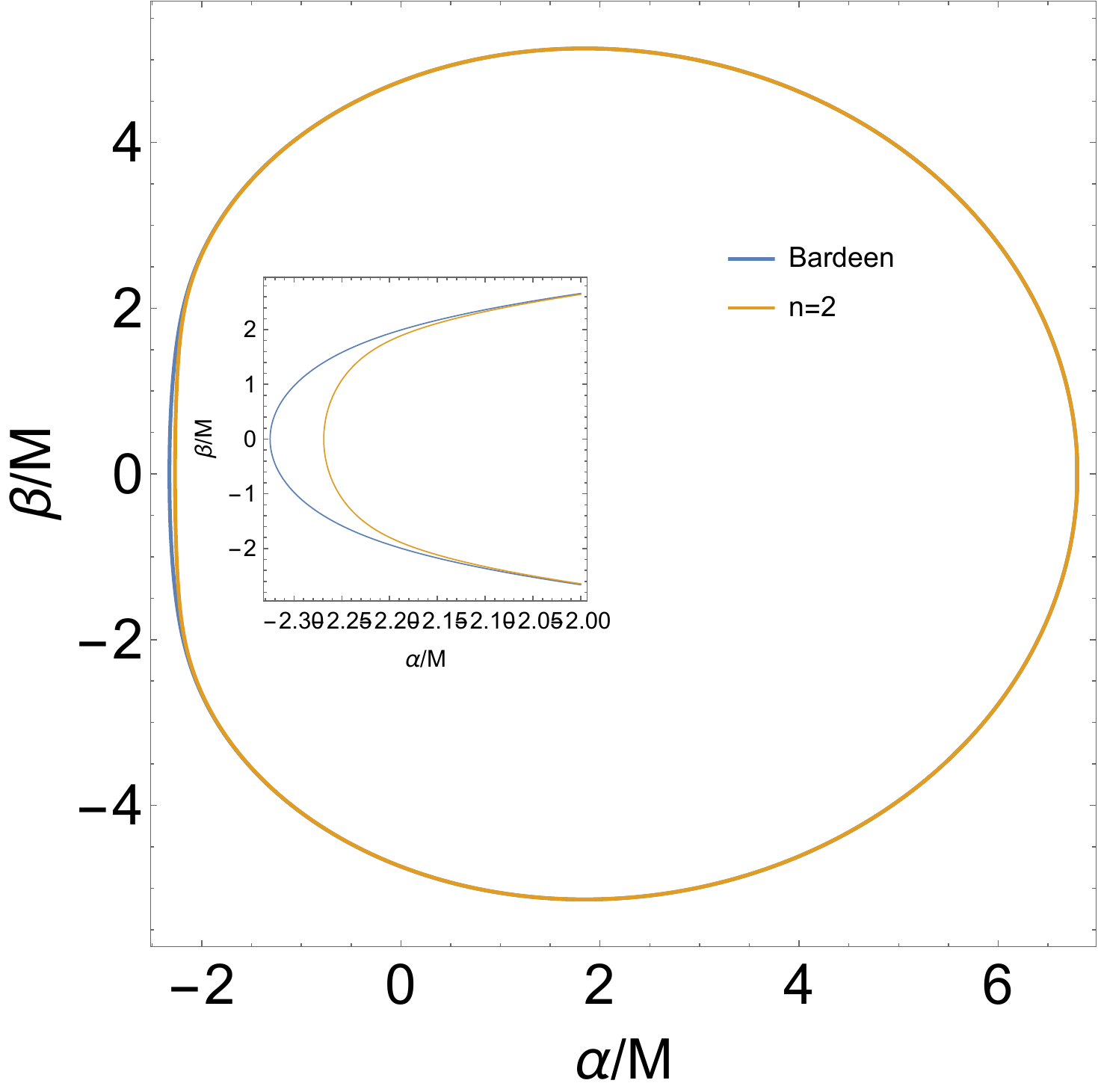}\ \hspace{0.05cm}
  \caption{The shadows of the black hole with $\gamma=2/3$ and $n=2$ and the Bardeen black hole with $\theta_0=\pi/2$. For both black holes, the parameters are fixed as $a/M=0.6$ and $g /\left(\sqrt{\frac{3}{2}} M^{2 / 3}\right)=0.64$ in the left plot , while  $a/M=0.9$ and $g /\left(\sqrt{\frac{3}{2}} M^{2 / 3}\right)=0.32$ in the right plot. The insets zoom in the part near the left edge of the circle.}\label{fig_2}
  }
\end{figure}

It is instructive to compare the differences of the shadow cast by  two different types of black holes, namely the Bardeen black hole and the regular black hole with $\gamma=2/3$ and $n=2$. Thus, we plot the shadows of these two black holes with the same values of parameters, as illustrated in Fig.\ref{fig_2}. Interestingly, we find the silhouette of shadow cast by the black hole with $\gamma=2/3$ and $n=2$ is more deformed than that of Bardeen black hole, although in general their sizes and shapes are quite similar. From the insets in Fig.\ref{fig_2}, one finds that they have distinct trajectory on the left edge of the circle. Such a discrepancy may be understood as the reflection of their different structure inside the horizon of black holes, where the former has a Minkowski core, while the latter has a de Sitter core. In particular, in \cite {Ling:2021olm} it is disclosed that for regular black hole with $\gamma=2/3$ and $n=2$, the position with the maximum of Kretschmann scalar curvature runs away from the core of the black hole, while for Bardeen black hole, the position of the maximal Kretschmann scalar curvature maintains at the center of the hole. In this sense,  the regular black hole with $\gamma=2/3$ and $n=2$ may become more attractive to photons such that the silhouette of shadow
is more deformed in comparison with that of Bardeen black hole.

\begin{figure} [th]
  \center{
  \includegraphics[scale=0.5]{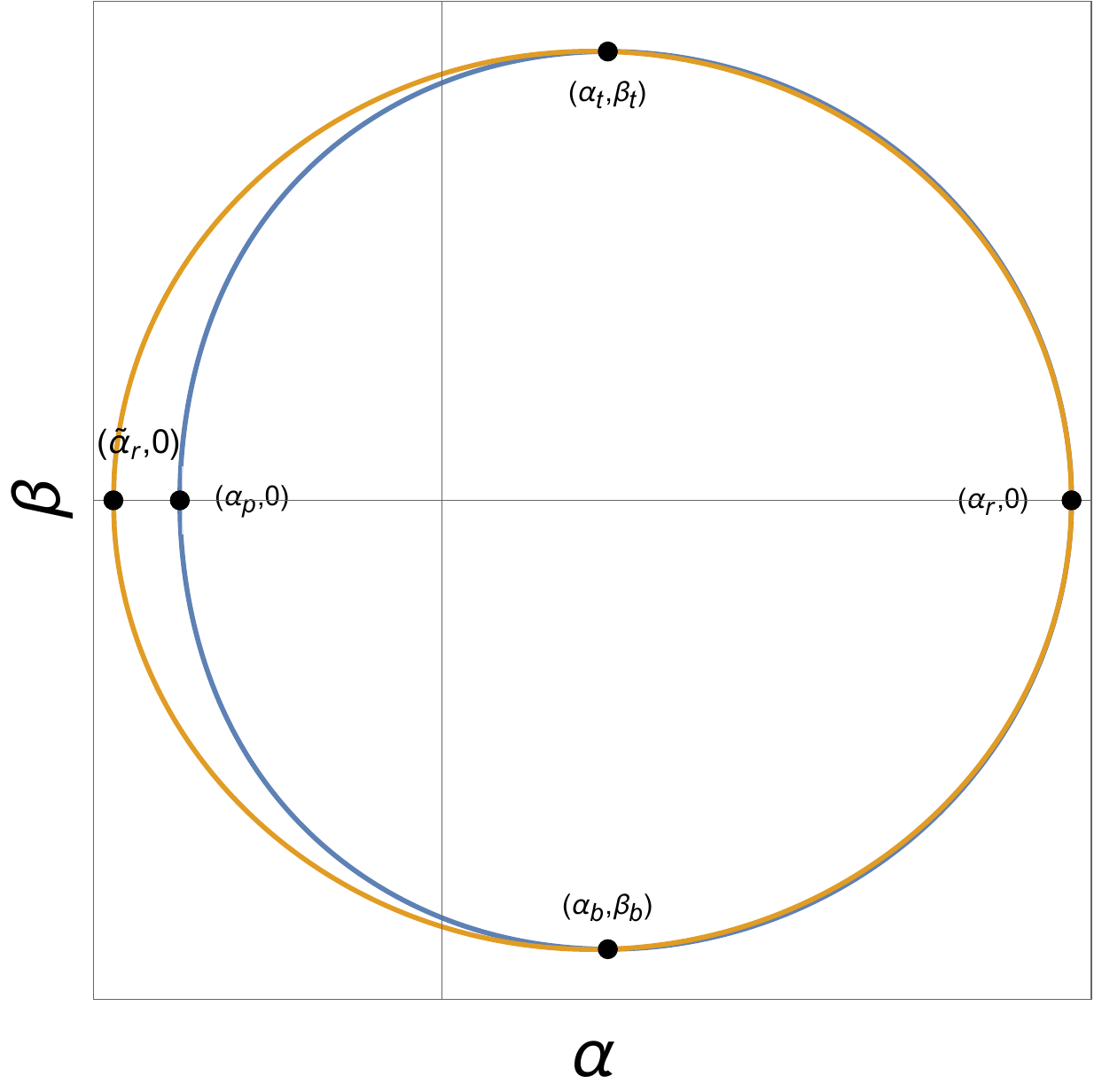}\ \hspace{0.05cm}
  \caption{The schematic diagram for defining the observables to measure the size and deformation of the shadow.  }\label{fig_3}
  }
\end{figure}

\begin{figure} [th]
  \center{
  \includegraphics[scale=0.315]{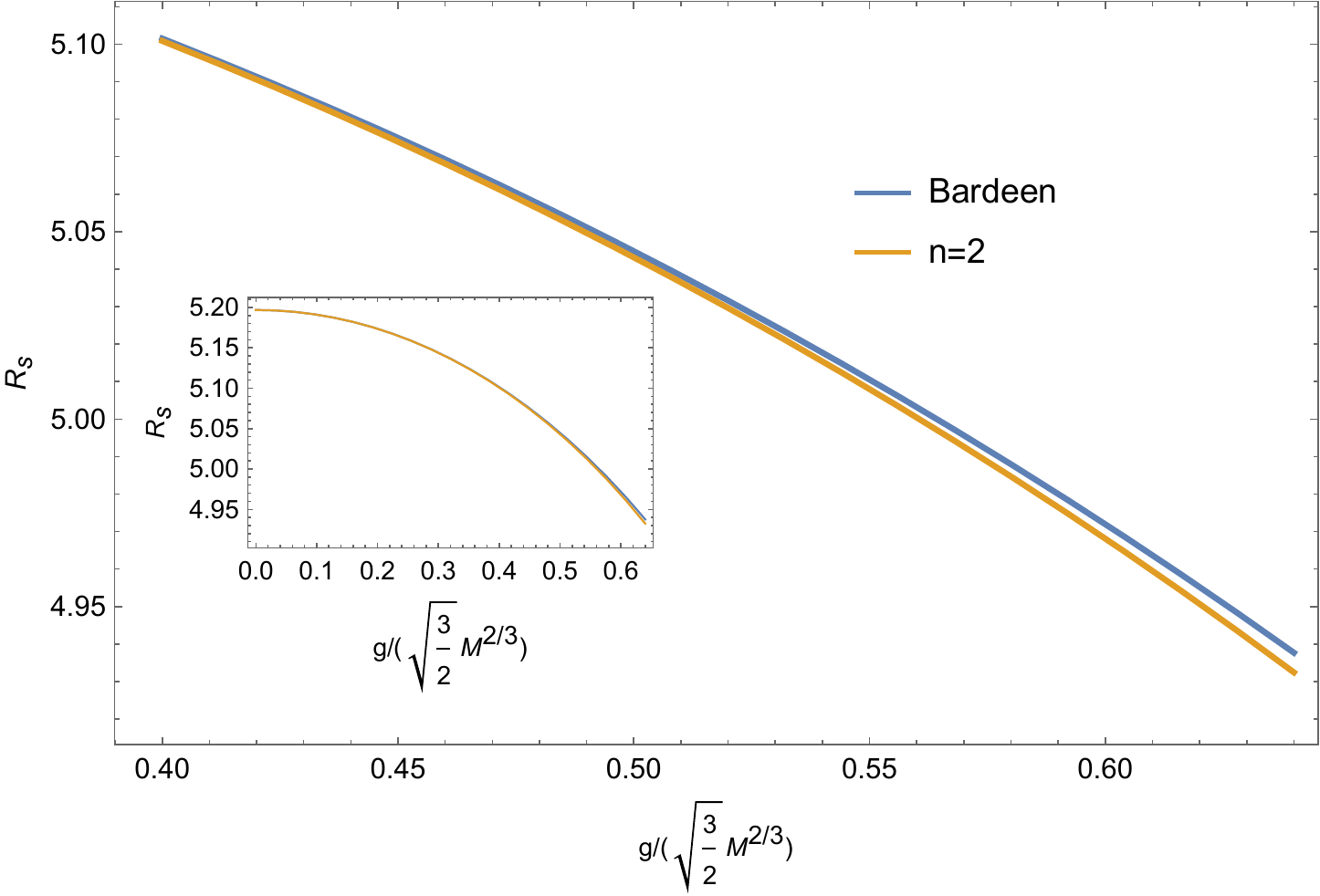}\ \hspace{0.05cm}
  \includegraphics[scale=0.33]{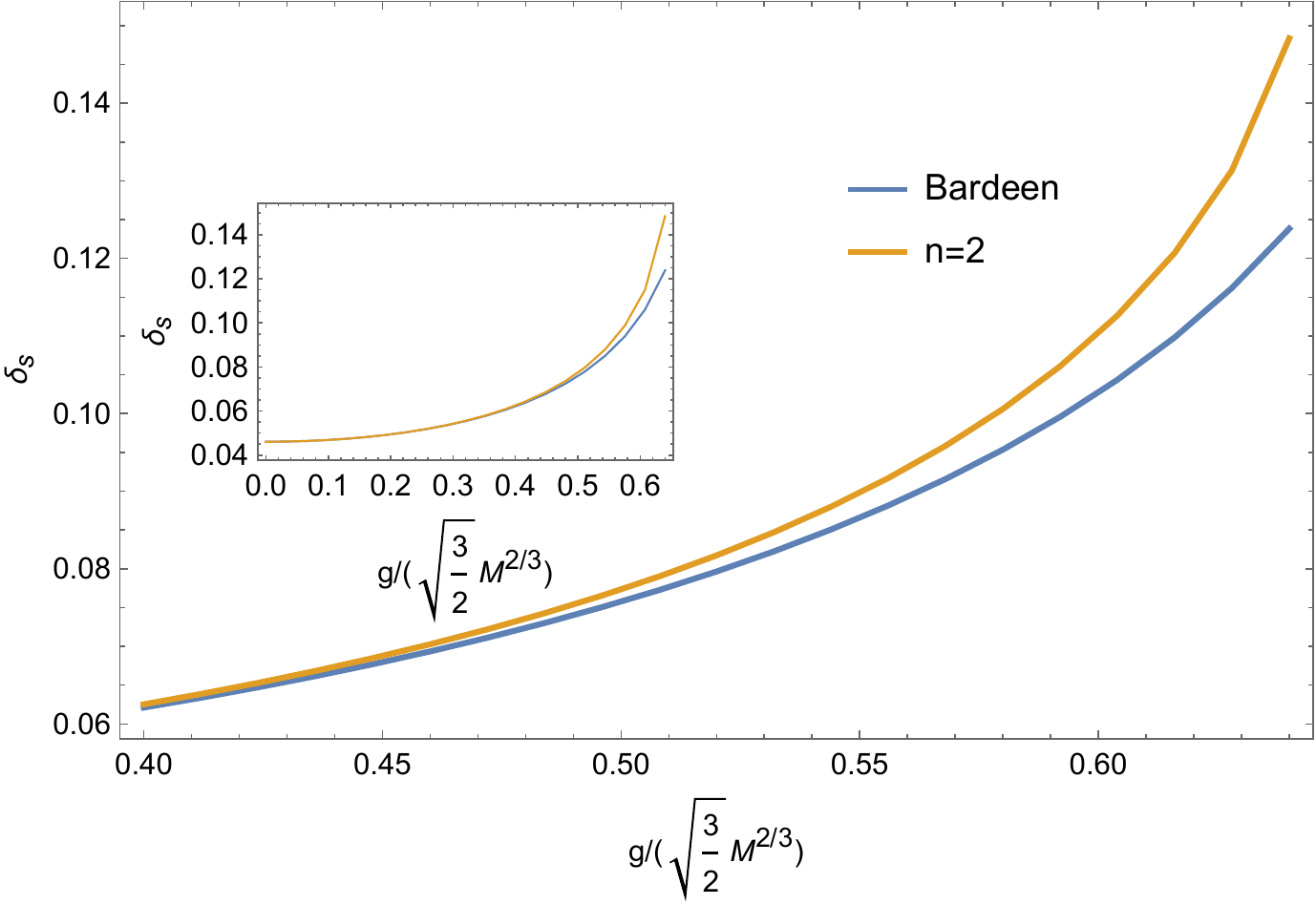}\ \hspace{0.05cm}
  \caption{ The shadow radius $R_s$ as the functions of  $g /\left(\sqrt{\frac{3}{2}} M^{2 / 3}\right)$(left). The distorting parameter $\delta_s$ as the functions of  $g /\left(\sqrt{\frac{3}{2}} M^{2 / 3}\right)$(right). The rotation parameter $a/M$ is fixed as $a/M=0.6$,  and  $\theta_0$ is fixed as $\pi/2$.}\label{fig_4}
  }
\end{figure}

Furthermore, in order to measure the size and deformation of the shadow quantitatively, we introduce two observables to characterize the shadow of black holes, namely the shadow radius $R_s$ and the distortion parameter $\delta_s$\cite{Hioki:2009na}, which are given by:

\begin{equation}
\begin{aligned}
R_{s}&=\frac{\left(\alpha_{t}-\alpha_{r}\right)^{2}+\beta_{t}^{2}}{2\left(\alpha_{t}-\alpha_{r}\right)}, \\
\delta_{s}&=\frac{\left(\tilde{\alpha}_{r}-\alpha_{p}\right)}{R_{s}},
\end{aligned}
\end{equation}
where $\left(\alpha_{t}, \beta_{t}\right),\left(\alpha_{r}, 0\right),\left(\alpha_{p}, 0\right)$ are the coordinates of the shadow vertices at top, right and left edges, while $\left(\tilde{\alpha}_{r}, 0 \right)$ is the coordinates of the left edges of the reference circle. The schematic diagram is shown in Fig.\ref{fig_3}. In Fig.\ref{fig_4}, we find that the shadow radius $R_s$ is monotonically decreasing as the function of  $g /\left(\sqrt{\frac{3}{2}} M^{2 / 3}\right)$ for both sorts of regular black holes, while the distortion parameter $\delta_s$ is monotonically increasing with the increase of the  deviation parameter. In particular, we find for larger values of the deviation parameter, the difference between two black holes becomes more pronounced, and the shadow cast by regular black hole with $\gamma=2/3$ and $n=2$ has more distortion than that cast by Bardeen black hole.

\begin{figure} [th]
  \center{
  \includegraphics[scale=0.9]{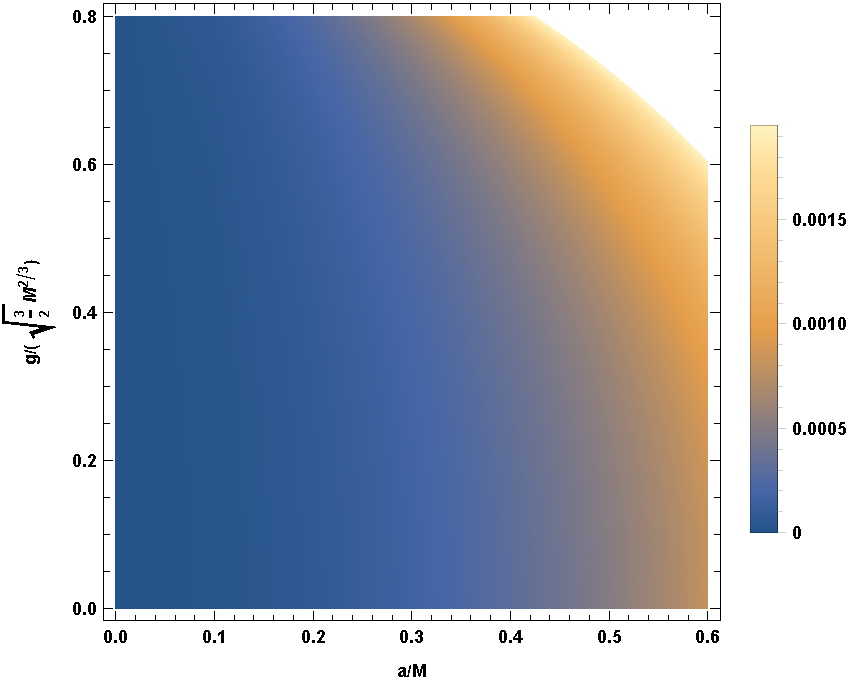}\ \hspace{0.05cm}
  \caption{The density plot of the deviation from circularity for the regular black hole with  $\gamma=1$ and $n=2$ in $(a,g)$ plane with  $\theta_0=17\pi/180$.}\label{fig_17n2}
  }
\end{figure}

Finally, we intend to link our theoretical investigation on
the shadow of regular black holes to the observation data detected
by ETH. As the first step, we consider the upper limit of the
deviation from circularity for the shadow of black holes.
As reported in
\cite{EventHorizonTelescope:2019dse,EventHorizonTelescope:2019ths},
for the supermassive black hole M87* at the centre of the galaxy
M87*, the upper limit of the deviation from circularity is $\Delta
C \lesssim 0.1$ for
$\theta_0=17\pi/180$\cite{CraigWalker:2018vam,Kuang:2022ojj}.  The
average radius of the shadow  is \cite{Johannsen:2010ru}
\begin{equation}
\begin{aligned}
\bar{R} &=\frac{1}{2 \pi} \int_{0}^{2 \pi} R(\varphi) d \varphi, \\
\end{aligned}
\end{equation}
with
\begin{equation}
\begin{aligned}
R(\varphi) &=\sqrt{\left(\alpha-\alpha_{c}\right)^{2}+\left(\beta-\beta_{c}\right)^{2}}, \ \ \varphi \equiv \tan ^{-1}\left(\frac{\beta-\beta_{c}}{\alpha-\alpha_{c}}\right),
\end{aligned}
\end{equation}
where $\alpha_{c}=\frac{\left|\alpha_{r }+\alpha_{p }\right|}{2}$,  $\beta_{c}=\frac{\left|\beta_{t }+\beta_{b }\right|}{2}$. The deviation from circularity $\Delta C$ is given by\cite{Afrin:2021imp}
\begin{equation}
\Delta C=\frac{1}{\bar{R}} \sqrt{\frac{1}{2 \pi} \int_{0}^{2
\pi}(R(\varphi)-\bar{R})^{2} d \varphi}.
\end{equation}
As shown in Fig.\ref{fig_17n2}, we take a contour plot on $\Delta
C$ in the $(a,g)$ plane for $\theta_0=17\pi/180$. It is obviously
to see that the deviation is much smaller than the upper limit
observed by ETH and indicates that the shadow of regular black
hole with $\gamma=2/3$ and $n=2$ is well compatible with the
observation data of the ETH. It implies that the current
observation data is not capable to identify the detected one to be
an ordinary black hole with singularity or a regular black hole
without singularity.

\begin{figure} [th]
  \center{
  \includegraphics[scale=0.35]{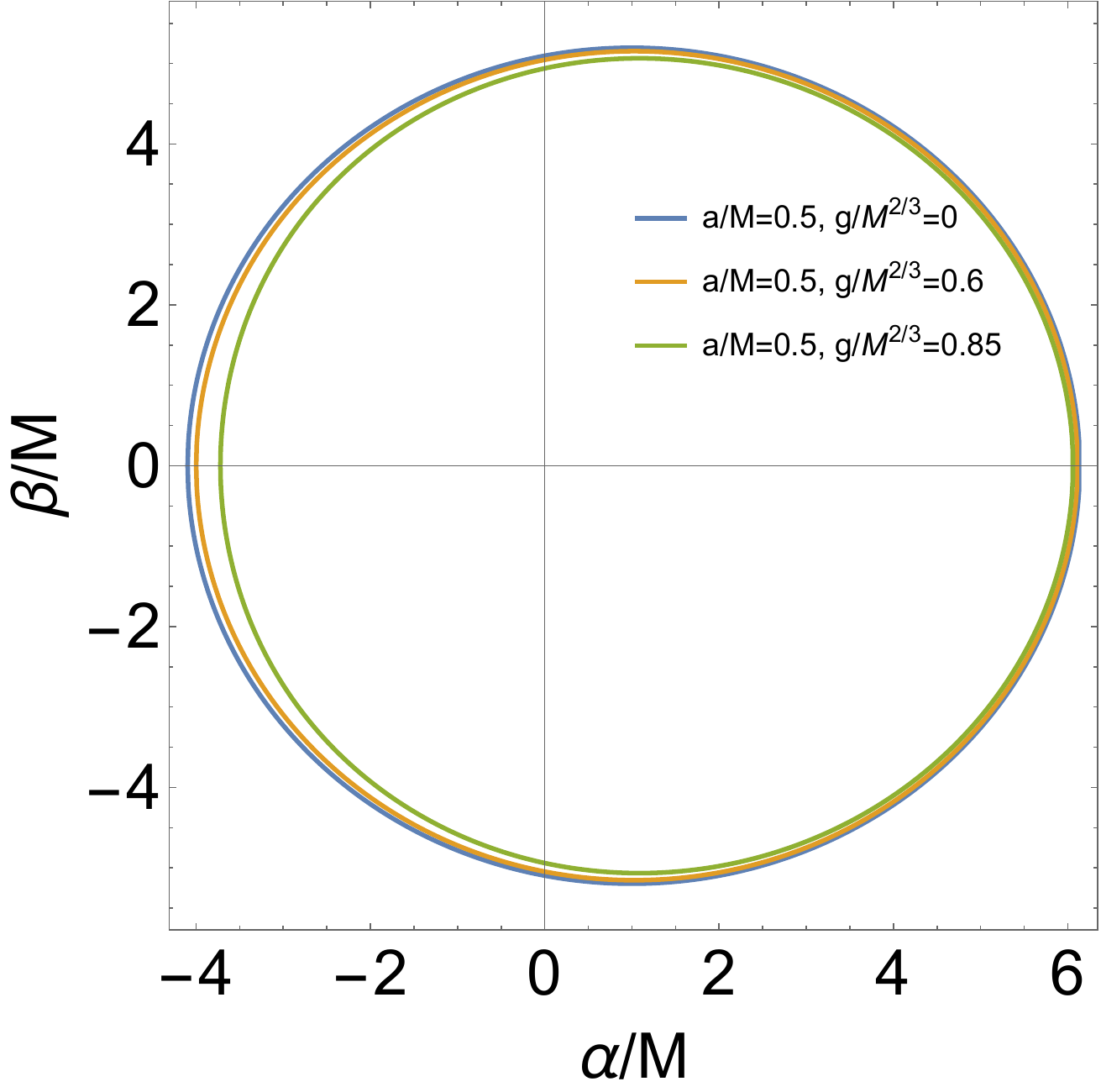}\ \hspace{0.05cm}
  \includegraphics[scale=0.34]{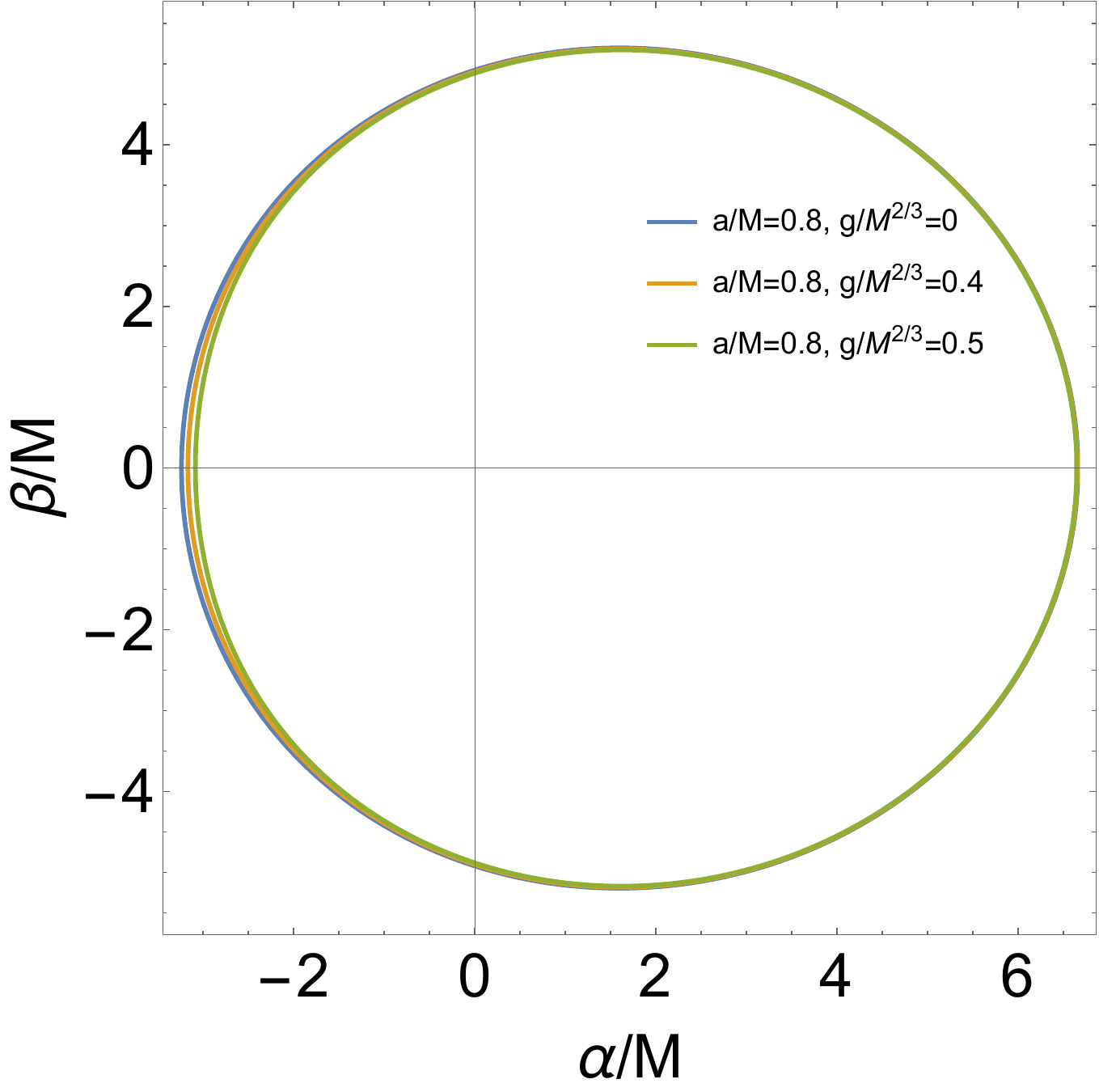}\ \hspace{0.05cm}
  \caption{The silhouette of the shadow cast by the regular black hole with  $\gamma=1$ and $n=3$ for various values of  $g /M^{2 / 3}$  with $\theta_0=\pi/2$. For the left plot, the rotation parameter $a/M$ is fixed as $a/M=0.5$ while for the right plot it is fixed as $a/M=0.8$.}\label{fig_5}
  }
\end{figure}

\section{Shadow of The Regular Black Hole with $\gamma=1$ and $n=3$ }

In this section we study the shadow of the regular black hole with $\gamma = 1$ and $n=3$ in a parallel way, which corresponds to Hayward black hole at large scale in radial direction.

We plot the silhouette of shadow for different values of $g /M^{2 / 3}$ in Fig.\ref{fig_5}. Again, we find  that with the increase of the values $g /M^{2 / 3}$, the size of the shadow shrinks, while the shape of the shadow becomes increasingly asymmetrical with respect to the vertical axis.

\begin{figure} [th]
  \center{
  \includegraphics[scale=0.3]{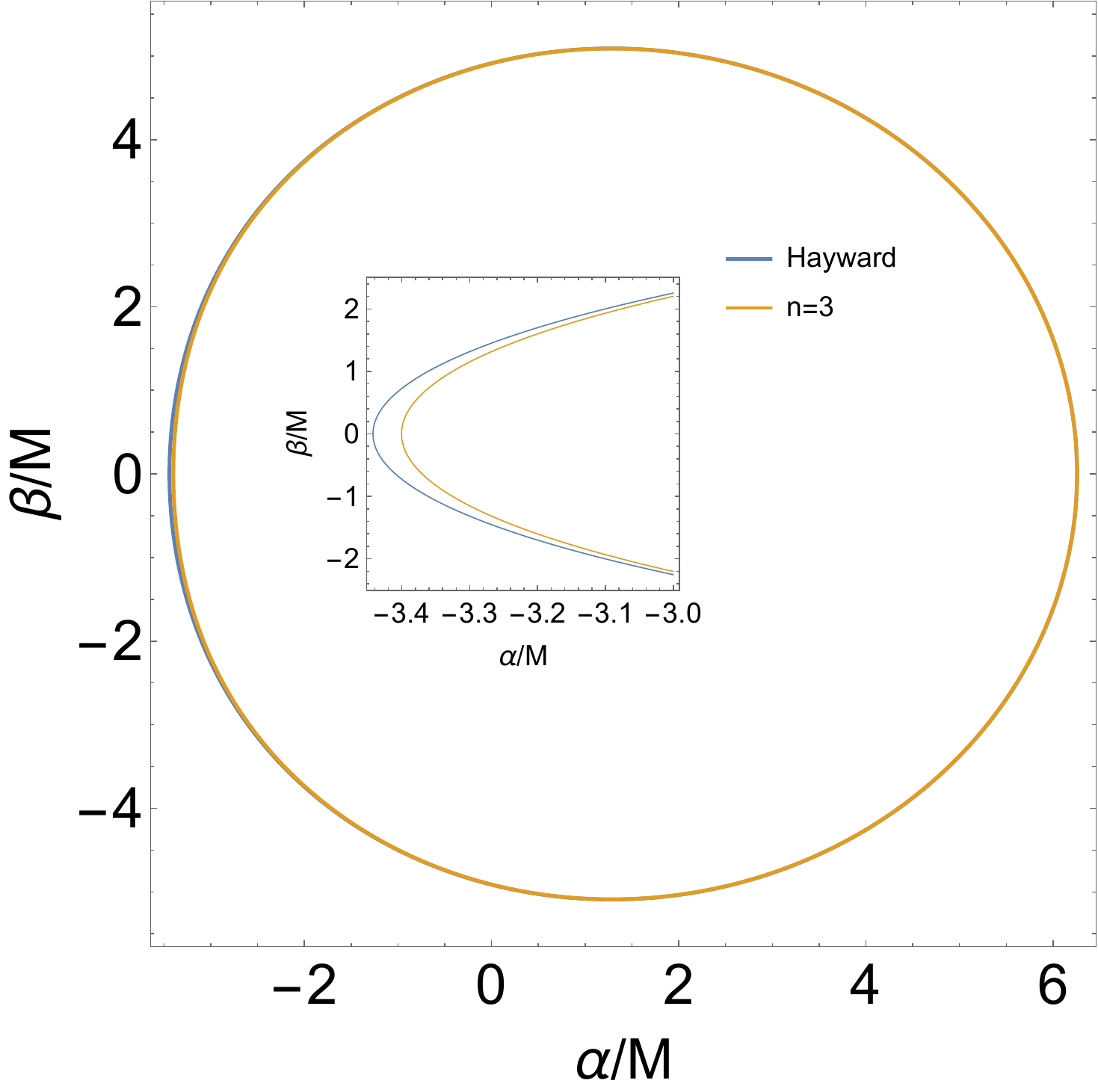}\ \hspace{0.05cm}
  \includegraphics[scale=0.3]{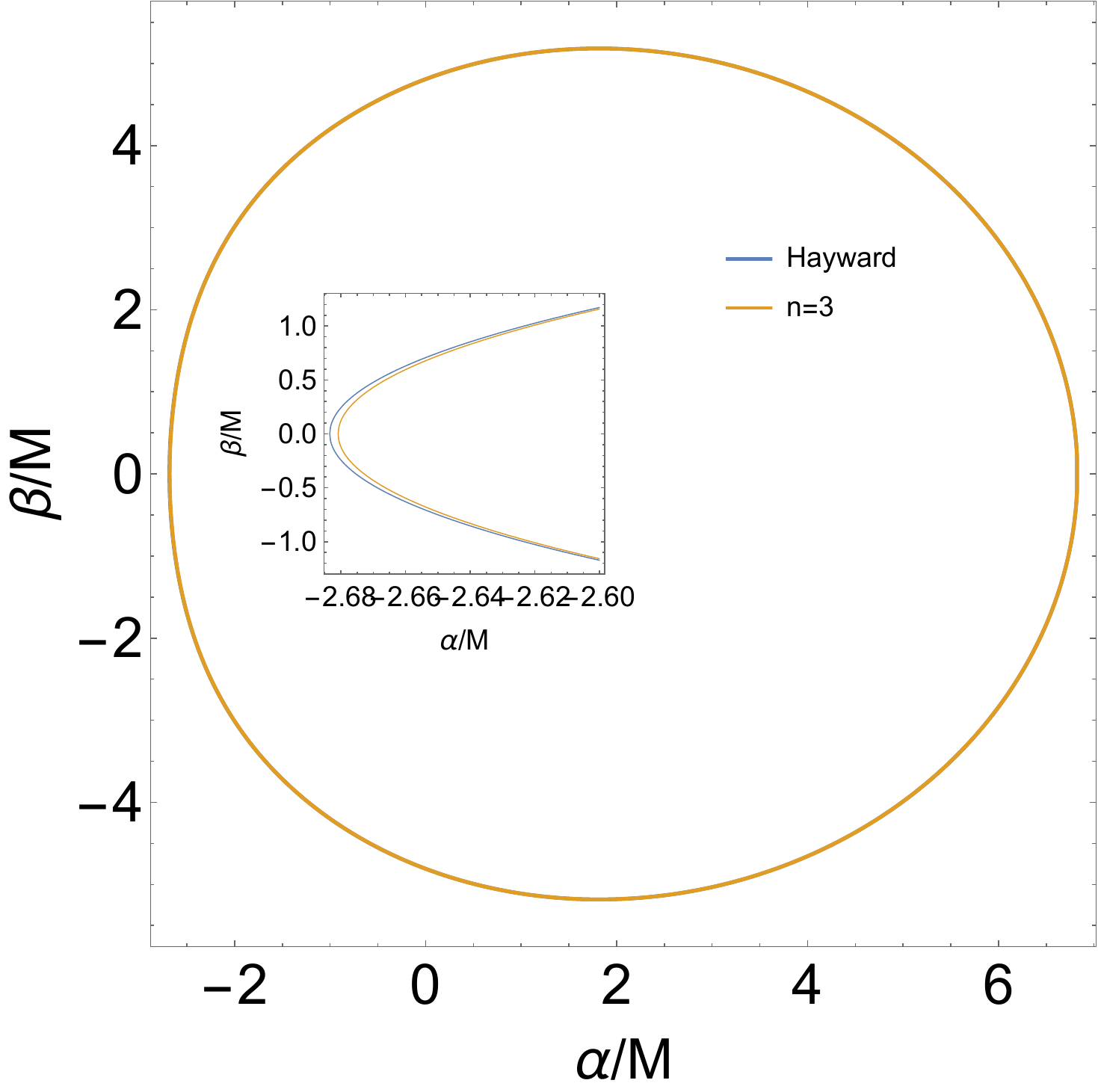}\ \hspace{0.05cm}
  \caption{The shadows of the black hole with $\gamma=1$ and $n=3$ and the Hayward black hole with $\theta_0=\pi/2$. For both black holes, the parameters are fixed as $a/M=0.6$ and $g /M^{2 / 3}=0.8$ in the left plot, while $a/M=0.9$ and $g /M^{2 / 3}=0.41$ in the right plot. The insets zoom in the part near the left edge of the circle.  }\label{fig_6}
  }
\end{figure}

\begin{figure} [th]
  \center{
  \includegraphics[scale=0.355]{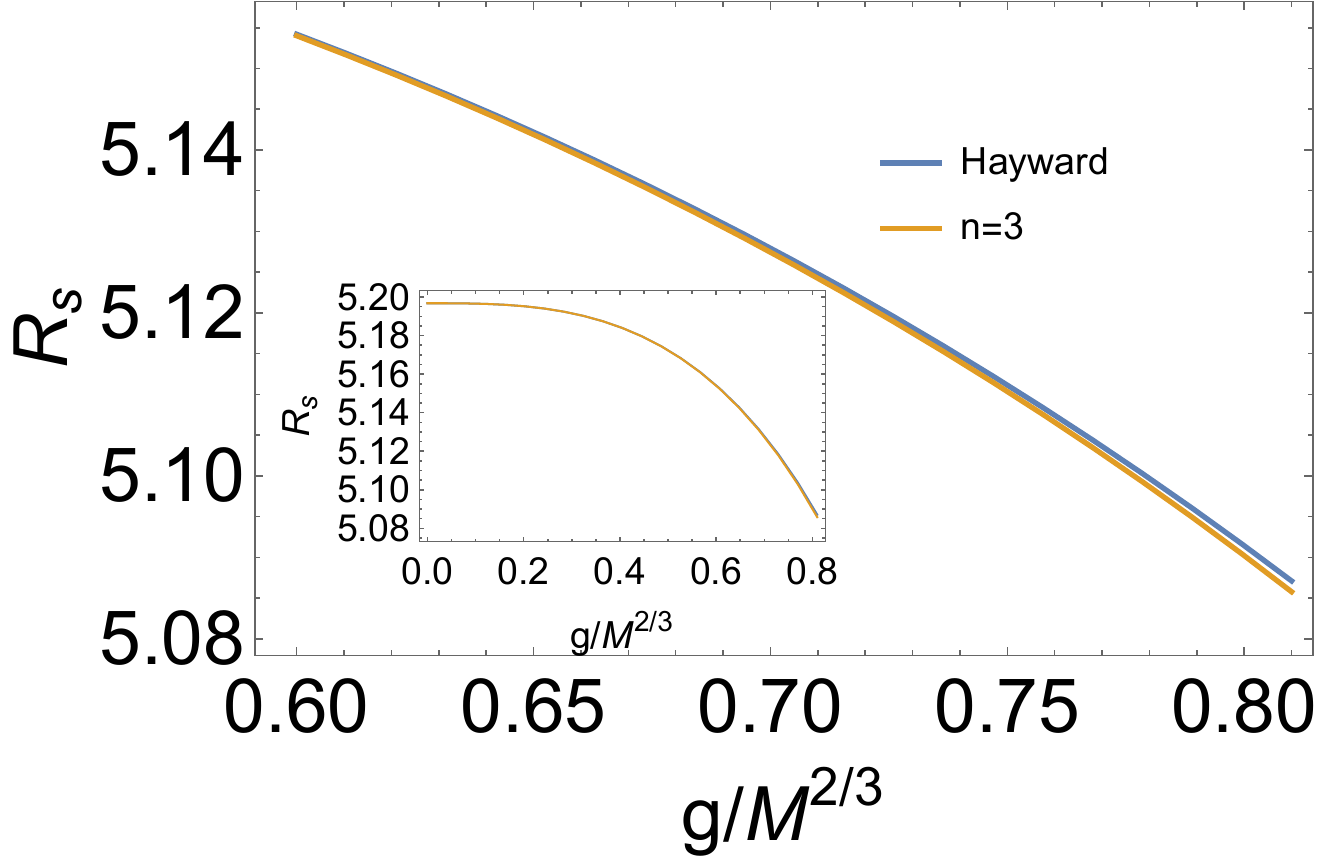}\ \hspace{0.05cm}
  \includegraphics[scale=0.345]{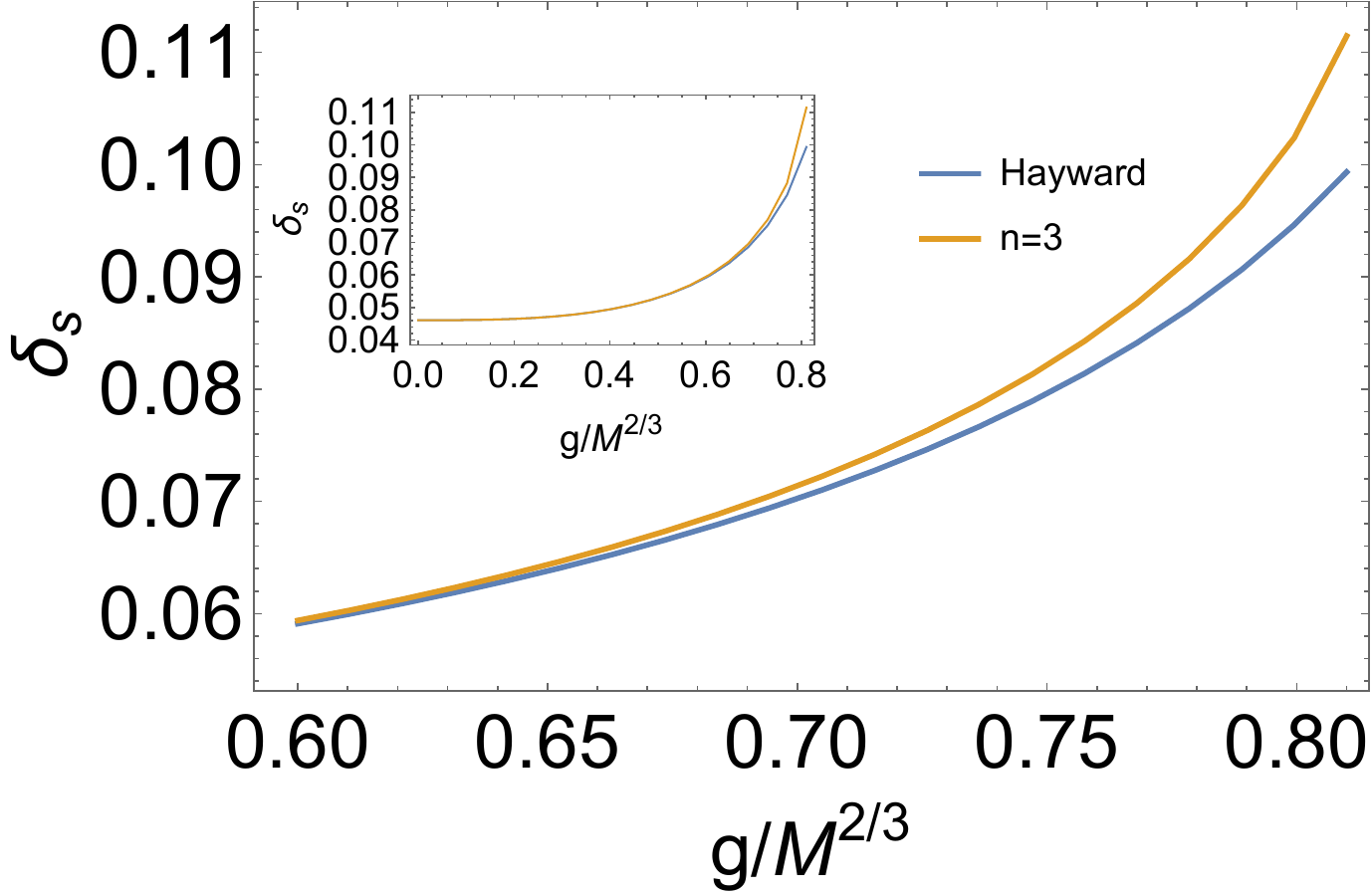}\ \hspace{0.05cm}
  \caption{The shadow radius $R_s$ as the functions of  $g /M^{2 / 3}$(left). The distorting parameter $\delta_s$ as the functions of $g /M^{2 / 3}$(right). The rotation parameter $a/M$ is fixed as $a/M=0.6$, and  $\theta_0$ is fixed as $\pi/2$.}\label{fig_7}
  }
\end{figure}

\begin{figure} [th]
  \center{
  \includegraphics[scale=0.9]{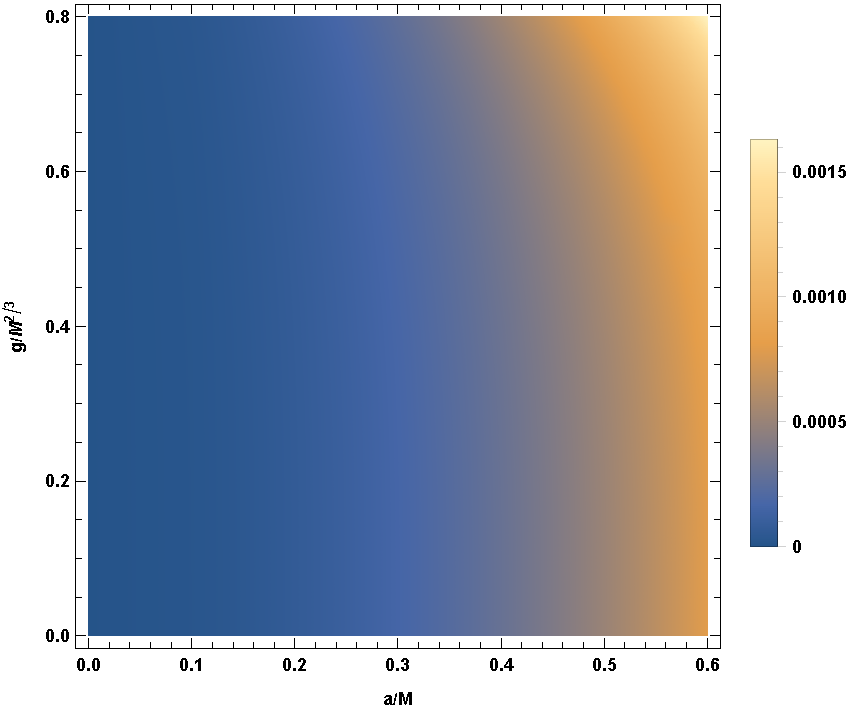}\ \hspace{0.05cm}
  \caption{The density plot of the deviation from circularity for the regular black hole with  $\gamma=1$ and $n=3$ in $(a,g)$ plane with  $\theta_0=17\pi/180$.}\label{fig_17n3}
  }
\end{figure}

Next we compare the differences between the shadow cast  by the regular black hole with $\gamma=1$ and $n=3$ and that by Hayward black hole, as shown in Fig.\ref{fig_6}. Similarly, we find that the size of the shadow cast by the regular black hole with $\gamma=1$ and $n=3$ is smaller than that of Hayward black hole with the same values of parameters. In addition, we plot diagrams for the shadow radius $R_s$ and the distortion parameter $\delta_s$ for these two black holes in Fig.\ref{fig_7}. Again, we notice that the difference between these two types of black holes becomes more significant for larger values of $g /M^{2/3}$. Therefore, based on our above analysis, we intend to conclude that given the same values of parameters, the size of the shadow cast by the regular black hole with asymptotically Minkowski core is always smaller than that of the black hole with asymptotically de Sitter core, but the deformation of the shadow is greater.

In the end, we also compare the theoretical value of
deviation from circularity $\Delta C $ for the black hole with
$\gamma=1$ and $n=3$ to the results of ETH, as shown in
Fig.\ref{fig_17n3}. We find that the shadow of regular black hole
with $\gamma=1$ and $n=3$ is well compatible with the observation
data of ETH $\Delta C \lesssim 0.1$ for $\theta_0=17\pi/180$.

\section{Conclusion and Discussion}

In this paper, we have investigated the shadow cast by rotating
Kerr-like regular black holes with Minkowski core. We have plotted
the silhouette of shadow cast by the regular black hole with
$(\gamma=2/3, n=2)$ and $(\gamma=1, n=3)$, which correspond to the
Bardeen black hole and Hayward black hole at large scales in
radial direction, respectively. It is found that with the increase
of the deviation parameter $g$, the left side of the silhouette of
shadow is more inclined to the vertical axis. Then, the size of
the shadow and the shape of the shadow have been evaluated by the
shadow radius and the distortion parameter. It turns out that with
the increase of the deviation parameter $g$, the shadow radius
decreases while the deformation gets more and more pronounced.
This phenomenon could be understood as follows. The
 traditional regular black holes can be viewed as solutions
of gravity coupled to nonlinear electrodynamics\cite{Ayon-Beato:1999qin,Ayon-Beato:1999kuh,Ma:2015gpa}, and the
parameter $g$ in this paper may be viewed as the charge of the
nonlinear electromagnetic field. Just as the increase of charge
makes the size of black hole horizon smaller, the
corresponding size of the shadow shrinks while the shape of
the shadow becomes more deformed\cite{Banerjee:2022iok,Kumar:2019pjp,Jafarzade:2020ova}.
Furthermore, we have compared the shadow for two different types
of regular black hole with the same values of parameters. In
comparison with the regular black holes with de Sitter core, the
size of the shadow is always smaller, and the deformation of the
shape becomes more pronounced as well.
 One could understand this by comparing the
position of the maximal Kretschmann scalar curvature. As revealed
in \cite{Ling:2021olm}, for regular black holes with Minkowski core, the position
of the maximal Kretschmann scalar curvature moves away from
the center of the black hole and is close to the horizon with the
increase of the parameter $g$, while  for regular black
holes with de Sitter core, the position of the maximal Kretschmann
scalar curvature always remains at the center. This may lead to
greater attraction of the regular black hole with Minkowski
core to photons. As a result, the silhouette of its shadow
is more distorted compared to that of the regular black hole with
de Sitter core indeed.
Finally, we have also
compared the theoretical values of the deviation from circularity
$\Delta C $ for two sorts of black holes with the experimental
results of the ETH and found that the shadow of regular black
holes is well compatible with the observed shadow by ETH as well.
Therefore, regular black holes are also possible candidates for
detected black holes and further exploration is needed to reveal
the nature of black holes observed in astrophysics. Our work
provides a theoretical basis for identifying different types of
regular black holes once more astronomical black hole images are
obtained in future.
Just as discussed in \cite{Ling:2021olm}, these two different types of
regular black holes results from different forms of Newtonian
potential, which might imply the different forms of generalized
uncertainty relations in the theory of quantum gravity. Thus we
expect that the distinction of regular black holes by astronomical
observations would be a great hint for us to look for the effect
of quantum gravity. In particular, the exponential form of
Newtonian potential, which gives rise to the regular black holes
with a Minkowski core\cite{Ling:2021olm}, could reveal the non-perturbative effects
of quantum gravity because in the expansion of weak momentum it
recovers the quadratic form of the momentum, which lead to the
regular black holes with a de-Sitter core\cite{Li:2016yfd}.

Of course we may investigate the shadow cast by regular black
holes in the presence of plasma as performed for Bardeen and
Hayward black holes in \cite{Abdujabbarov:2016hnw}. It is worth
pointing out that what we have analyzed in this paper is an
integrable system due to the existence of the Carter constant. It
is interesting to consider a non-integrable system surrounding the
black hole by breaking the Carter constant\cite{Wang:2021gja}. In
this case, the trajectory of the light near the black hole may
produce chaos, which will provide more information for the regular
black holes.

\section*{Acknowledgments}

We are very grateful to Prof. Xin Wu, Prof. Xiaomei Kuang and Dr. Hong Guo for helpful discussions.
This work is supported in part by the Natural Science Foundation
of China under Grant No.~11875053 and 12035016. It is also supported by Beijing Natural Science Foundation under Grant No. 1222031.

\end{document}